\documentclass[conference]{IEEEtran}
\IEEEoverridecommandlockouts
\usepackage{cite}
\usepackage{amsmath,amssymb,amsfonts}
\usepackage[linesnumbered,ruled,vlined]{algorithm2e}
\usepackage{algorithmicx}
\usepackage{algpseudocode}
\usepackage{multirow}
 \algdef{SE}[DOWHILE]{Do}{doWhile}{\algorithmicdo}[1]{\algorithmicwhile\ #1}%
\usepackage{graphicx}
\def\BibTeX{{\rm B\kern-.05em{\sc i\kern-.025em b}\kern-.08em T\kern-.1667em\lower.7ex\hbox{E}\kern-.125emX}}
\usepackage{calrsfs}
\DeclareMathAlphabet{\pazocal}{OMS}{zplm}{m}{n}
\usepackage{textcomp}
\usepackage{xcolor}
\usepackage[export]{adjustbox}
\graphicspath{{Schematics/}}

\renewenvironment{thebibliography}[1]{%
	\begin{oldthebibliography}{#1}%
		\setlength{\itemsep}{-0.05cm}%
	}%
	{
	\end{oldthebibliography}
}
\def\BibTeX{{\rm B\kern-.05em{\sc i\kern-.025em b}\kern-.08em
    T\kern-.1667em\lower.7ex\hbox{E}\kern-.125emX}}
\begin{document}

\title{Visible light communication-based monitoring for indoor environments using unsupervised learning}

\author{\IEEEauthorblockN{Mehmet C. Ilter$^{1}$, Alexis A. Dowhuszko$^{2}$, Jyri H\"am\"al\"ainen$^{2}$ and Risto Wichman$^{1}$}  
\IEEEauthorblockA{
$^{1}$Department of Signal Processing and Acoustics, Aalto University, 02150 Espoo, Finland\\
$^{2}$Department of Communications and Networking, Aalto University, 02150 Espoo, Finland\\
Email: \{mehmet.ilter, alexis.dowhuszko, jyri.hamalainen, risto.wichman\}@aalto.fi}}

\maketitle

\begin{abstract}
Visible Light Communication~(VLC) systems provide not only illumination and data communication, but also indoor monitoring services if the effect that different events create on the received optical signal is properly tracked. For this purpose, the Channel State Information that a VLC receiver computes to equalize the subcarriers of the OFDM signal can be also reused to train an Unsupervised Learning classifier. This way, different clusters can be created on the collected CSI data, which could be then mapped into relevant events to-be-monitored in the indoor environments, such as the presence of a new object in a given position or the change of the position of a given object. When compared to supervised learning algorithms, the proposed approach does not need to add tags in the training data, simplifying notably the implementation of the machine learning classifier. The practical validation the monitoring approach was done with the aid of a software-defined VLC link based on OFDM, in which a copy of the intensity modulated signal coming from a Phosphor-converted LED was captured by a pair of Photodetectors~(PDs). The performance evaluation of the experimental VLC-based monitoring demo achieved a positioning accuracy in the few-centimeter-range, without the necessity of deploying a large number of sensors and/or adding a VLC-enabled sensor on the object to-be-tracked.

\end{abstract}

\vspace{2mm}

\begin{IEEEkeywords}
Visible Light Communications; Phosphor-Converted LED; Silicon PIN diode; Indoor monitoring; Optical OFDM; Unsupervised Learning.
\end{IEEEkeywords}

\section{Introduction}
\label{sec:1}

Activity recognition has gained notable attention lately thanks to the large number of applications that should monitor the movement of humans to carry out their tasks. Traditionally, activity recognition has relied on the information collected from wearable sensors or cameras that continuously take readings or pictures~\cite{chan2008}. However, these methods have drawbacks related to the energy consumption, infrastructure costs and/or users’ privacy concerns. This is the reason why the utilization of wireless communication signal for activity recognition represents an interesting option, as the infrastructure for sensing is already deployed and the users’ privacy can be protected~\cite{zafa2019}.

Visible Light Communication (VLC) technology tackles most of the drawbacks that monitoring systems based on Radio Frequency (RF) communications have~\cite{karu2015, path2015}. For example, since VLC beams cannot propagate through opaque obstacles such as walls or curtains, they can be easily confined into the designated coverage area to enable secured communication links. In addition, though VLC uses a portion of the electromagnetic spectrum that is licence-free, it enables the ultra-dense deployment of co-located low-interference beams using directive spot-lights~\cite{guzman2020}. Thanks to this, the stability of the communication link increases, and it is possible to detect minor changes on the received optical signal, which enables a better activity recognition accuracy when compared to RF-based solutions. Moreover, the same infrastructure for illumination and communication can be also re-used for VLC-based monitoring. Finally, VLC technology can be utilized in places such as hospitals, where RF systems are banned due to Electromagnetic Compatibility problems.

Most of State-of-the-Art~(SotA) solutions for VLC-based monitoring require a sensor on the object to-be-tracked. Then, by measuring the optical power that reaches the Photodetector (PD), it is possible to extract relevant metrics to estimate the location of the object. Examples of these metrics are \emph{e.g.} Angle-of-Arrival~(AoA), Received Signal Strength~(RSS), Time-of-Arrival~(ToA), and Time-Difference-of-Arrival~(TDoA)~\cite{rahman2020recent}. Recently, novel machine learning solutions have been proposed to extract useful patterns from the VLC channel for monitoring purposes. For example, the authors of~\cite{ilter2020a,ilter2020b} presented an object identification and object localization algorithm, respectively, using a Random Forest Classifier that was trained using the Channel State Information~(CSI) that was collected in presence of different target events. However, the main drawback of such supervised learning algorithms is the data to train the classifier must be properly \emph{labeled}, which is a requirement that complicates notably the implementation of these algorithms in practice.  

In order to tackle this limitation, in this paper we implement an \emph{unsupervised learning algorithm} that identifies the most convenient number of clusters in which the training data should be divided. Since each of these clusters will be associated to a different target event to-be-monitored, the necessity of labeled training data is avoided, and the partitioning of the training data is created without the necessity of human intervention. For the practical validation of this unsupervised learning approach, a software-defined VLC link using Universal Software Radio Peripherals~(USRPs), low-cost phosphor-converted LEDs~\cite{luxe2017}, and commercial Photodetectors~\cite{thor2019} was utilized. This way, the Supervised Learning classifier was trained with actual CSI used to detect the transmitted data stream, without affecting the performance of the VLC link.

The  rest  of  the  paper  is  organized  as  follows:  Section~\ref{sec:2} describes the implemented software-defined VLC link and explains the effect that different obstacles have on the received optical spectrum at different positions, which represent the unique signature that we aim at identifying using the CSI measurements. Section~\ref{sec:3} addresses the key concepts behind the proposed Supervised Learning algorithm, and adapts it to the task that is needed. Then, Section~\ref{sec:4} explains the experimental setting that was used to validate the VLC-based monitoring concept, and carries out the performance evaluation when sample objects takes different positions on the service area. Finally, conclusions are drawn in Section~\ref{sec:5}.

\bibliographystyle{IEEEtran}
\bibliography{VTCArxiv}
\end{document}